\documentclass[%
 aip,
 amsmath,amssymb,
 reprint,
]{revtex4-2}

\usepackage{graphicx}
\usepackage{dcolumn}
\usepackage{bm}
\usepackage[utf8]{inputenc}
\usepackage[T1]{fontenc}
\usepackage{mathptmx}
\usepackage{etoolbox}
\usepackage{hyperref}
\usepackage{gensymb}
\usepackage{xcolor}
\usepackage{soul}    
\usepackage{comment}

\makeatletter
\def\@email#1#2{%
 \endgroup
 \patchcmd{\titleblock@produce}
  {\frontmatter@RRAPformat}
  {\frontmatter@RRAPformat{\produce@RRAP{*#1\href{mailto:#2}{#2}}}\frontmatter@RRAPformat}
  {}{}
}%
\makeatother

\begin{document}

\preprint{AIP/123-QED}

\title[]{Mapping the High-Field Response of Superconducting Nanowire Single-Photon Detectors }

\author{Harshit Agarwal}
\affiliation{Materials Science and Technology Division, Oak Ridge National Laboratory, Oak Ridge, Tennessee 37831, USA}

\author{Yueh-Chun Wu}
\affiliation{Materials Science and Technology Division, Oak Ridge National Laboratory, Oak Ridge, Tennessee 37831, USA}

\author{Katyayani Seal}
\affiliation{Single Quantum B.V., Delft, the Netherlands}

\author{Niels Noordzij}
\affiliation{Single Quantum B.V., Delft, the Netherlands}

\author{Andreas Fognini}
\affiliation{Single Quantum B.V., Delft, the Netherlands}

\author{Ronan Gourgues}
\affiliation{Single Quantum B.V., Delft, the Netherlands}

\author{G\'abor B. Hal\'asz}
\affiliation{Materials Science and Technology Division, Oak Ridge National Laboratory, Oak Ridge, Tennessee 37831, USA}

\author{Benjamin J. Lawrie}
\email{lawriebj@ornl.gov}
\affiliation{Materials Science and Technology Division, Oak Ridge National Laboratory, Oak Ridge, Tennessee 37831, USA}

\author{Nirjhar Sarkar}
\email{nirjharize@gmail.com}
\affiliation{Materials Science and Technology Division, Oak Ridge National Laboratory, Oak Ridge, Tennessee 37831, USA}

\date{\today}

\begin{abstract}
Superconducting nanowire single-photon detectors (SNSPDs) are increasingly being integrated into cryogenic quantum-sensing platforms that operate in large magnetic fields. Here, the high-field limits of these measurements were probed by characterizing the dark count response of a NbTiN SNSPD at up to 14~T using in situ control of the detector orientation and benchmarking that performance against normal detector operation in smaller magnetic fields. The detector performance is strongly anisotropic: perpendicular magnetic fields rapidly reduce the usable bias-current range, whereas near-parallel alignment enables operation at substantially larger total fields, with a finite operating margin retained to 12~T and measurable dark-count response at the 14~T field limit. The angular dependence is governed predominantly by the perpendicular component of the magnetic field, consistent with vortex-mediated suppression of the superconducting current scale. The same field-dependent current scale also determines the electrical pulse amplitude, linking high-field detector operation directly to readout performance. These results establish a practical route for deploying SNSPDs in multi-tesla quantum-sensing and high-field cryogenic experiments.\footnote{This manuscript has been authored by UT-Battelle, LLC, under contract DE-AC05-00OR22725 with the US Department of Energy (DOE). The US government retains and the publisher, by accepting the article for publication, acknowledges that the US government retains a nonexclusive, paid-up, irrevocable, worldwide license to publish or reproduce the published form of this manuscript, or allow others to do so, for US government purposes. DOE will provide public access to these results of federally sponsored research in accordance with the DOE Public Access Plan (http://energy.gov/downloads/doe-public-access-plan).}
\end{abstract}

\maketitle

%\section{\label{sec:intro}Introduction}

Superconducting nanowire single-photon detectors (SNSPDs) were first patterned from ultrathin NbN films in 2001~\cite{goltsman2001}. In the twenty-five years since, SNSPDs have become the detector of choice for quantum optics and photonics because of their near-unity detection efficiency, picosecond timing jitter, and sub-hertz dark-count rates~\cite{natarajan2012,holzman2019superconducting,venza2025research}. Increasingly, however, SNSPDs are being incorporated into cryogenic experiments that also require substantial magnetic fields. These include integrated quantum-photonic platforms involving spin transitions and trapped ions~\cite{awschalom2018,bruzewicz2019}, as well as emerging applications in dark-matter detection, nuclear and high-energy physics, and high-field quantum sensing~\cite{chiles2022,ahmed2018quantum,abdul2022science,solanki2026subterahertz}. Reliable SNSPD operation in large magnetic fields is therefore becoming an important requirement for a growing range of quantum-sensing platforms.

Past studies of the magnetic-field-dependent response of SNSPDs have shown that the detector response depends strongly on the field orientation: perpendicular fields rapidly suppress the accessible bias-current range and modify the dark-count response, whereas substantially larger fields can be tolerated when the field is aligned close to the device plane~\cite{engel2012dependence,lusche2014,polakovic2020,lawrie2021}. Exploiting this geometry, SNSPD operation has previously been demonstrated in near-parallel fields of approximately 5--6 T~\cite{polakovic2020,lawrie2021}. However, detector performance at substantially higher fields needed for high energy and nuclear physics applications has not yet been experimentally established~\cite{chiles2022,ahmed2018quantum,polakovic2020}.

Here, we characterize the dark-count response of an SNSPD in fields of up to 14~T using in situ control of the detector orientation. The dark-count-limited bias current range is mapped as a function of field amplitude and orientation, yielding an estimated several-microampere bias-current margin at 12~T relative to the photon-count onset current measured at low perpendicular fields and a measurable dark count response at the 14~T field limit. At low perpendicular fields, the photon-count onset current varies very weakly with field compared with the dark-count threshold currents, which shift rapidly toward lower bias current. The field-dependent SNSPD response is further correlated with the measured voltage-pulse amplitude, which is strongly suppressed in increasing fields.

\begin{figure*}[htb]
\centering
\includegraphics[width=\textwidth]{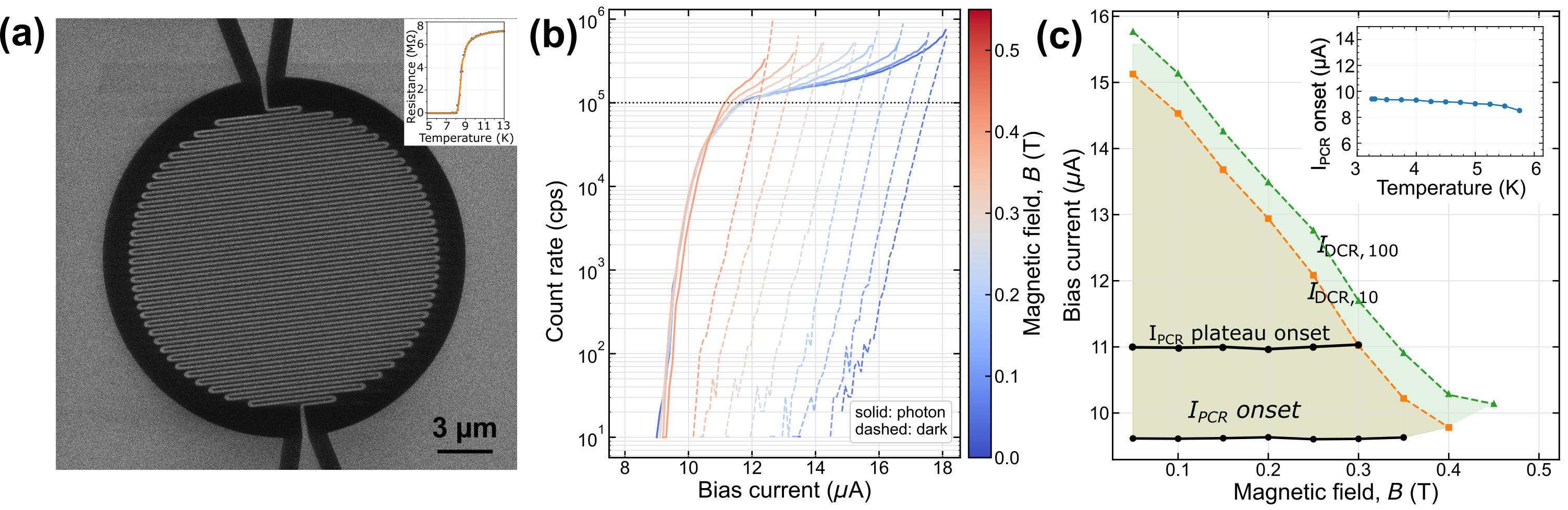}
\caption{\label{fig:device_dcr}
Device characterization and perpendicular-field response of the SNSPD.
(a) Helium ion micrograph of the circular meandered NbTiN SNSPD. The inset shows the resistance vs temperature of the device.
(b) Photon-count (solid) and dark-count (dashed) rates as a function of bias current for increasing perpendicular magnetic field at $T=5$~K under 500~nm wavelength illumination.
(c) Extracted bias current $I_{\mathrm{PCR}}$ associated with the onset of measured photon counts and the onset of the photon-count-rate plateau (described in further detail in the Supplementary Information), along with the dark-count threshold currents $I_{\mathrm{DCR},10}$, and $I_{\mathrm{DCR},100}$ as a function of perpendicular magnetic field. Shaded regions indicate the corresponding bias-current windows. The inset shows the $I_{\mathrm{PCR}}$ onset current as a function of temperature with zero field.
}
\end{figure*}

The NbTiN SNSPD investigated in this work is shown in Fig.~\ref{fig:device_dcr}(a). The device consists of a 10-nm-thick, 100-nm-wide nanowire patterned with a 200-nm pitch on a telecom-optimized distributed Bragg reflector. The resistance measurement shown in the inset of Fig.~\ref{fig:device_dcr}(a) gives a superconducting transition that begins near 8~K. High-field dark-count measurements up to 14~T were performed in a Quantum Design DynaCool PPMS equipped with a 14~T superconducting magnet. An Attocube piezoelectric rotator integrated into a custom PPMS probe allowed in situ control of the detector orientation. Photon-count measurements under 500~nm wavelength continuous-wave illumination were performed separately in a Quantum Design OptiCool cryostat equipped with a 7~T superconducting magnet. These measurements were performed with perpendicular magnetic fields up to approximately 0.5~T. Further details of the sample positioning and magnetic-field specifications are provided in Supplementary Information, Sec.~S1. Detector biasing and pulse readout were performed using a Single Quantum ATLAS SNSPD driver/controller using a 100 mV threshold for all detection events. The angle $\theta$ is defined such that $\theta=0^\circ$ corresponds to the field parallel to the detector plane and $\theta=90^\circ$ to the perpendicular configuration, with $B_{\perp}=B\sin\theta$.

The detector response is first characterized with the magnetic field applied perpendicular to the device plane. Figure~\ref{fig:device_dcr}(b) shows the photon- and dark-count rates as a function of bias current for increasing perpendicular magnetic field measured at $T=5 K$. The photon-count rate is obtained by subtracting the dark-count rate from the total count rate measured under illumination. The photon-count onset current varies only weakly over the measured field range, whereas the dark-count curves shift progressively toward lower bias current. This contrasting response is consistent with the distinct physical origins of the two processes reported previously~\cite{engel2012dependence,lusche2014,vodolazov2015}. The small shift in the photon-count onset current at the highest field may reflect changes in current distribution or vortex dynamics, although experimental variability cannot be excluded. The shift of the dark-count threshold currents toward lower bias current is consistent with the field-dependent vortex-entry barrier: as $B_{\perp}$ lowers this barrier and suppresses the critical-current scale, a given activation rate is reached at a smaller absolute bias current. Photon detection is instead governed primarily by the photon-induced hotspot, making the photon-count onset current substantially less sensitive to the same field range~\cite{engel2012dependence,lusche2014,vodolazov2015}.

\begin{figure*}[htb!]
\centering
\includegraphics[width=\textwidth]{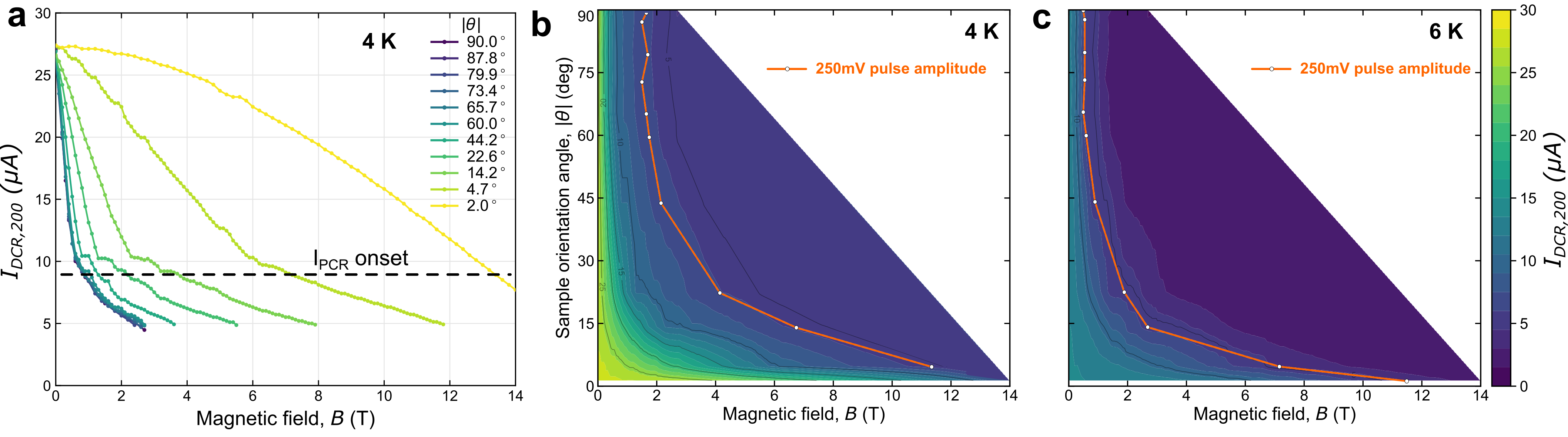}
\caption{\label{fig:isw_angle}
Magnetic-field amplitude and orientation dependence of the dark-count-defined current scale.
(a) $I_{\mathrm{DCR},200}$ as a function of magnetic field at 4~K for different detector orientations.
(b,c) Interpolated maps of $I_{\mathrm{DCR},200}$ as a function of magnetic field and detector orientation at 4 and 6~K, respectively. The color scale represents $I_{\mathrm{DCR},200}$, and the black curves denote constant-$I_{\mathrm{DCR},200}$ contours. The orange contour denotes the measured 250~mV pulse-amplitude discussed further in Fig.~\ref{fig:pulse_metrics}.
}
\end{figure*}

The corresponding operating window is quantified in Fig.~\ref{fig:device_dcr}(c). The photon-count onset current, $I_{\mathrm{PCR}}$, is defined as the bias current at which the logarithmic slope $d\log_{10}(R_{\mathrm{PCR}})/dI_{\mathrm{bias}}$ reaches its maximum, where $R_{\mathrm{PCR}}$ is the photon-count rate. The photon-count plateau boundary, where the plateau begins, is identified at the subsequent minimum of this slope (Supplementary Information, Sec.~S3 and Fig.~S2). The dark-count threshold current, $I_{\mathrm{DCR},n}$, is defined as the bias current at which the dark-count rate reaches $n$~cps. Figure~\ref{fig:device_dcr}(c) shows $I_{\mathrm{DCR},10}$ and $I_{\mathrm{DCR},100}$, which exhibit similar field dependence and shift toward lower bias currents with increasing $B_{\perp}$. As $B_{\perp}$ increases, the dark-count threshold currents move rapidly toward $I_{\mathrm{PCR}}$, while $I_{\mathrm{PCR}}$ changes much more weakly, progressively narrowing the usable bias-current window. The inset further shows that $I_{\mathrm{PCR}}$ remains nearly constant over the measured temperature range from 3 to 6~K. This low-perpendicular-field behavior provides the baseline for the high-field angle-dependent measurements discussed below.

The rapid narrowing of the operating window in perpendicular field motivates measurements with in-plane fields where substantially larger total fields can be tolerated. The high-field response was therefore measured as a function of both magnetic-field magnitude and detector orientation at temperatures of 4 and 6~K. To quantify the upper operating boundary consistently across these measurements, $I_{\mathrm{DCR},200}$ is defined as the bias current at which the dark-count rate reaches 200~cps. Figure~\ref{fig:isw_angle}(a) shows $I_{\mathrm{DCR},200}$ as a function of magnetic field at 4~K for different detector orientations. A strong angular dependence is observed: configurations with a larger perpendicular field component lose accessible bias current rapidly, whereas near-parallel alignment preserves a measurable current scale to much larger total fields. Taking the $\sim9~\mu$A photon-count onset current from Fig.~\ref{fig:device_dcr}(c) as a reference, the closest-to-parallel trace ($|\theta|=2.0^\circ$, highlighted in yellow in Fig.~\ref{fig:isw_angle}(a)) retains a finite bias-current margin of order a few microamperes at 12~T (assuming similarly weak field dependence at near-parallel orientations, as observed under perpendicular fields in Fig.~\ref{fig:device_dcr}(c)). A measurable $I_{\mathrm{DCR},200}$ persists to the 14~T field limit, although by this field it falls below the photon-count onset current.

A distinct change in slope is also evident in most of the $I_{\mathrm{DCR},200}(B)$ traces in Fig.~\ref{fig:isw_angle}(a), occurring at different total magnetic fields for different orientations. Similar behavior has been reported in NbN meanders, where the critical current decreases approximately linearly with field in the Meissner regime before entering a flatter mixed-state behavior following vortex penetration~\cite{lusche2014}. Higher-field steps and kinks in such devices have further been associated with vortex rearrangement and the nucleation of additional vortex rows, while pinning and meander geometry can modify the subsequent field dependence~\cite{lusche2014,vodolazov2015}. The feature observed here is therefore consistent with the onset or reorganization of a vortex-containing state, although $I_{\mathrm{DCR},200}$ is a dark-count-defined current scale rather than a direct measurement of the thermodynamic critical current.

\begin{figure*}[htb]
\centering
\includegraphics[width=\textwidth]{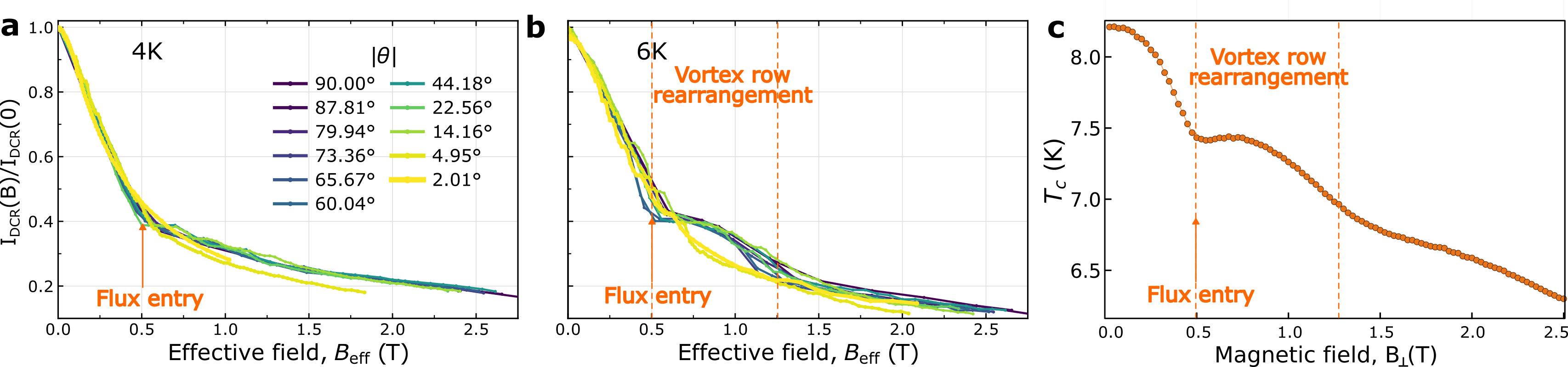}
\caption{\label{fig:B_per}
Effective-field scaling of the normalized dark-count threshold current $I_{\mathrm{DCR},200}$ as a function of $B_{\mathrm{eff}}=B_{\perp}/(1-\alpha B^2)$ for different detector orientations at (a) 4~K and (b) 6~K, using $\alpha=0.00312~\mathrm{T}^{-2}$.
(c) Superconducting transition temperature $T_c$ as a function of perpendicular magnetic field.}
\end{figure*}

Current--voltage measurements at 4~K show that the switching (critical) current $I_c$ decreases systematically with increasing perpendicular magnetic field from 0 to 3~T, as shown in Supplementary Fig.~S1(a). The measured $I_c$ also exhibits an approximately symmetric dependence on perpendicular magnetic field [Supplementary Fig.~S1(c)]. This behavior is consistent with the opposite orientations of the hairpin bends, where field-induced screening currents enhance current crowding at one set of bends while reducing it at the other, with these effects reversed upon field reversal~\cite{engel2012dependence}.

The individual angular traces in Fig.~\ref{fig:isw_angle}(a) are combined into the two-dimensional field--orientation map in Fig.~\ref{fig:isw_angle}(b). Each constant-angle trace corresponds to a horizontal cross section at fixed $\theta$, with the color scale giving $I_{\mathrm{DCR},200}(B,\theta)$. The corresponding map at 6~K is shown in Fig.~\ref{fig:isw_angle}(c) as an example of the limiting behavior as the temperature approaches $T_c$. The overall current scale is reduced at the higher temperature, while the strong dependence on detector orientation remains.

The angular dependence is examined further in Fig.~\ref{fig:B_per} using a phenomenological effective field $B_{\mathrm{eff}}=B_{\perp}/(1-\alpha B^2)$, where $B_{\perp}=B\sin\theta$ and $\alpha$ accounts empirically for additional field-induced suppression of superconductivity. A single fitted value of $\alpha=0.00312~\mathrm{T}^{-2}$ improves the collapse of the normalized current traces at both 4 and 6~K. The dominant perpendicular-field dependence remains consistent with vortex-mediated suppression, with a characteristic crossover near $B_{\perp}\sim0.55$~T, where the intervortex spacing $a\sim\sqrt{\Phi_0/B_{\perp}}\approx60$--$70$~nm is comparable to the 100-nm nanowire width where $\Phi_0=h/(2e)$ is the superconducting magnetic flux quantum. At 6~K, an additional weak nonmonotonic feature is visible above the initial crossover. Similar nonmonotonic field dependencies have been associated with changes in confined vortex-row configurations and vortex rearrangement in narrow superconducting strips~\cite{mcnaughton2022causes,vodolazov2013vortex,morgan2015measurement}. The measured $T_c$ also decreases with perpendicular magnetic field, exhibiting a weak shoulder near 0.5~T [Fig.~\ref{fig:B_per}(c)], close to the crossover observed in the current response and $I_c$ (Supplementary Fig.~S1(a)). This feature may be associated with vortex reconfiguration~\cite{chibotaru2000symmetry} .Its weaker appearance at 4~K is likely due to reduced vortex mobility and the stronger edge-barrier or pinning-limited behavior expected at lower temperature~\cite{engel2008temperature}. The closest-to-parallel trace ($\theta=2.0^\circ$) shows the largest deviation from the simple perpendicular-field scaling. At this orientation, the larger total magnetic field required for a given $B_{\perp}$ may introduce additional magnetic pair breaking and modify the vortex-entry barrier, potentially masking the characteristic crossover ~\cite{xi2010far,lee2023gapless}. Small angular misalignments may also contribute to the observed deviation. At 14~T, an angular offset of $0.1^\circ$ corresponds to approximately 63~mT in $B_{\mathrm{eff}}$, as the near-parallel configuration is particularly sensitive to angular uncertainty.

\begin{figure*}[htb!]
\centering
\includegraphics[width=\textwidth]{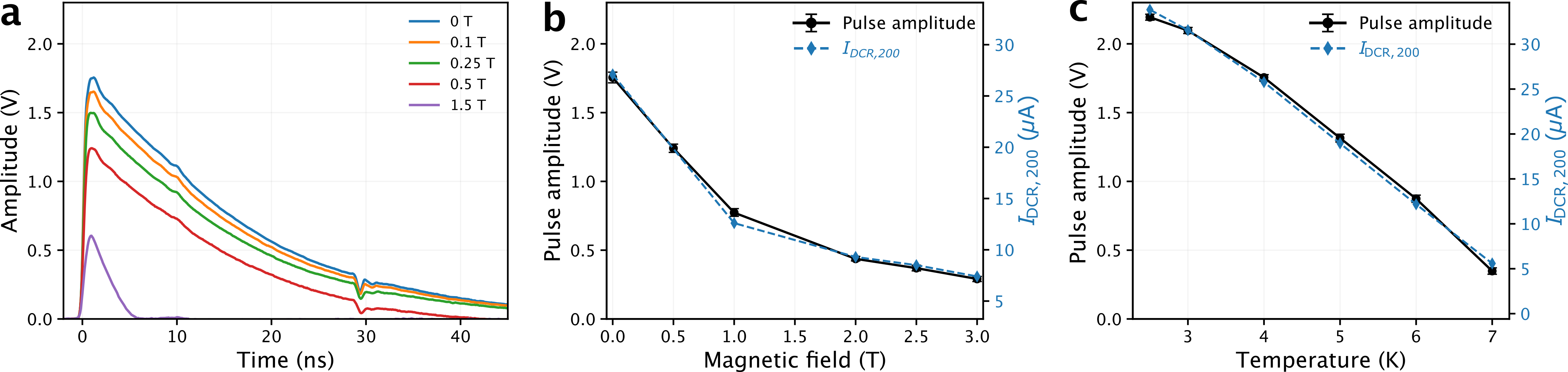}
\caption{\label{fig:pulse_metrics}
Magnetic-field- and temperature-dependent SNSPD pulse amplitude.
(a) Measured median voltage pulse waveforms at 4~K for different magnetic fields.
(b) Measured pulse amplitude as a function of magnetic field at 4~K, compared with the independently extracted $I_{\mathrm{DCR},200}$ on the right axis.
(c) Measured pulse amplitude as a function of temperature, compared with $I_{\mathrm{DCR},200}$ on the right axis.
}
\end{figure*}

%The constant-$I_{\mathrm{DCR},200}$ contours in Figs.~\ref{fig:isw_angle}(b) and \ref{fig:isw_angle}(c) provide a practical extension of the field--orientation map for applications where the signal-to-noise of the SNSPD pulse amplitudes are important in high field applications. 
In the standard SNSPD current-diversion model, the voltage-pulse amplitude is approximately~\cite{kerman2006kinetic}
\begin{equation}
V_{\mathrm{pulse}}\simeq (I_{\mathrm{bias}}-I_{\mathrm{ret}})Z_0G,
\end{equation}
where $I_{\mathrm{ret}}$ is the current remaining in the nanowire when the resistive region collapses during a detection pulse, $Z_0$ is the readout impedance, and $G$ is the total gain of our readout amplifiers. The observed reduction in pulse amplitude with increasing magnetic field is consistent with the decrease in the available SNSPD operating current. The orange contour in Figs.~\ref{fig:isw_angle}(b) and \ref{fig:isw_angle}(c) marks the field--orientation combinations corresponding to a 250~mV pulse amplitude, providing a representative pulse-amplitude criterion for the present readout. The relation between the bias current, temperature, magnetic field, and pulse amplitude is examined directly in Fig.~\ref{fig:pulse_metrics}. Figure~\ref{fig:pulse_metrics}(a) shows measured median voltage pulse waveforms at 4~K for increasing magnetic field. The pulse amplitude decreases systematically with increasing magnetic field. The pulse rise time increases while the fall time decreases with increasing magnetic field, as quantified in Supplementary Fig.~S3. The same dependence is evident in Fig.~\ref{fig:pulse_metrics}(b), where the pulse amplitude is compared with the independently extracted $I_{\mathrm{DCR,200}}$. Both quantities decrease together with increasing magnetic field, demonstrating a correlation between the dark-count threshold current and pulse amplitude.~\cite{kerman2006kinetic}. However, because the maximum out-of-plane field for normal operation is $\sim$0.3 T, as highlighted in Fig.~1, these changes do not have substantial implications for normal SNSPD operation.

A closely related behavior is observed under temperature tuning. Figure~\ref{fig:pulse_metrics}(c) shows that both $I_{\mathrm{DCR},200}$ and the pulse amplitude decrease with increasing temperature and closely follow the same temperature dependence. Temperature therefore affects not only the bias-current range over which the SNSPD can be operated, but also the electrical signal available to the readout. Together with the field dependence in Fig.~\ref{fig:pulse_metrics}(b), this correspondence provides the basis for using the constant-current contours in Figs.~\ref{fig:isw_angle}(b) and \ref{fig:isw_angle}(c) as approximate pulse-amplitude operating boundaries.

In summary, the high-field response of a NbTiN SNSPD was characterized using dark-count measurements, in-situ field-orientation control, and waveform analysis. Increasing perpendicular magnetic field rapidly narrows the usable bias-current window, while near-parallel alignment extends the operating range to substantially larger total fields, retaining a several-microampere operating window at 12~T and a measurable $I_{\mathrm{DCR},200}$ at the 14~T field limit. The resulting field--orientation maps provide a practical measure of the angular range available for maintaining a desired detector operating current at a given magnetic field. At 4~K, the angular dependence follows a common dependence on $B_{\perp}$, including the crossover near $B_{\perp}\sim0.55$~T, consistent with a response dominated by perpendicular-field vortex physics. The close correspondence between $I_{\mathrm{DCR},200}$ and the voltage-pulse amplitude under both temperature and magnetic-field tuning further connects the available operating current to the electrical signal delivered to the readout. These results thus establish a practical approach and provide concrete guidelines for operating and orienting SNSPDs in multi-tesla magnetic-field environments. Improved control over vortex pinning sites established through controlled growth, device design\cite{van2026geometry} and ion-beam patterning~\cite{strohauer2025current,sarkar2026understanding} could further extend the limits defined here to even higher field environments.

\begin{acknowledgments}
This research was sponsored by the U.\,S. Department of Energy, Office of Science, Basic Energy Sciences, Materials Sciences and Engineering Division. \end{acknowledgments}

\section*{Data Availability Statement}
The data that support the findings of this study will be posted to an openly shared repository.

\end{document}